\documentclass[iop,revtex4]{emulateapj}
\usepackage{graphicx}
\usepackage{amssymb}
\usepackage{amsmath}
\usepackage{threeparttable}
\usepackage{lscape}

\mathchardef\mhyphen="2D
\newcommand{\aliii}{Al\,{\sc iii}}

\newcommand{\siv}{S\,{\sc iv}}
\newcommand{\siiv}{Si\,{\sc iv}}

\newcommand{\pv}{P\,{\sc v}}

\newcommand{\civ}{C\,{\sc iv}}

\newcommand{\Feii}{Fe\,{\sc ii}}
\newcommand{\angstrom}{\text{ \normalfont\AA}}

\mathchardef\mhyphen="2D

\def\civ{C\,{\sc iv}}
\def\niii{N\,{\sc iii}}

\def\oiv{O\,{\sc iv}}

\def\siiv{Si\,{\sc iv}}
\def\Siii{Si\,{\sc ii}}

\def\siv{S\,{\sc iv}}

\def\aliii{Al\,{\sc iii}}
\def\pv{P\,{\sc v}}

\def\nh{\ifmmode n_\mathrm{\scriptstyle H} \else $n_\mathrm{\scriptstyle H}$\fi}
\def\ne{\ifmmode n_\mathrm{\scriptstyle e} \else $n_\mathrm{\scriptstyle e}$\fi}
\def\Qh{\ifmmode Q_\mathrm{\scriptstyle H} \else $Q_\mathrm{\scriptstyle H}$\fi}
\def\Uh{\ifmmode U_\mathrm{\scriptstyle H} \else $U_\mathrm{\scriptstyle H}$\fi}
\def\Nh{\ifmmode N_\mathrm{\scriptstyle H} \else $N_\mathrm{\scriptstyle H}$\fi}

\def\Zsun{\ifmmode {\rm Z}_{\odot} \else Z$_{\odot}$\fi}
\def\Msun{\ifmmode {\rm M}_{\odot} \else M$_{\odot}$\fi}
\def\kms{\ifmmode {\rm km~s}^{-1} \else km~s$^{-1}$\fi}
\def\Lya{\ifmmode {\rm Ly}\alpha \else Ly$\alpha$\fi}
\def\Lyb{\ifmmode {\rm Ly}\beta \else Ly$\beta$\fi}
\def\Lyg{\ifmmode {\rm Ly}\gamma \else Ly$\gamma$\fi}
\def\Lyd{\ifmmode {\rm Ly}\delta \else Ly$\delta$\fi}
\def\neaod{\ifmmode n_\mathrm{\scriptscriptstyle AOD} \else $n_\mathrm{\scriptscriptstyle AOD}$\fi}
\def\necrit{\ifmmode n_\mathrm{\scriptstyle cr} \else $n_\mathrm{\scriptstyle cr}$\fi}
\def\ncr{\ifmmode n_\mathrm{\scriptstyle cr} \else $n_\mathrm{\scriptstyle cr}$\fi}
\def\nepi{\ifmmode n_\mathrm{\scriptscriptstyle PI} \else $n_\mathrm{\scriptscriptstyle PI}$\fi}
\def\gtorder{\mathrel{\raise.3ex\hbox{$>$}\mkern-14mu\lower0.6ex\hbox{$\sim$}}}
\def\ltorder{\mathrel{\raise.3ex\hbox{$<$}\mkern-14mu\lower0.6ex\hbox{$\sim$}}}

\shorttitle{ }
\shortauthors{Xu et al.}
\shortauthors{}

\begin{document}


\title{VLT/X-Shooter Survey of BAL quasars: Large Distance Scale and AGN Feedback}


\author{
Xinfeng Xu\altaffilmark{1,$\dagger$},
Nahum Arav\altaffilmark{1},
Timothy Miller\altaffilmark{1},
Chris Benn\altaffilmark{2}
}

\affil{$^1$Department of Physics, Virginia Tech, Blacksburg, VA 24061, USA}
\affil{$^2$Isaac Newton Group, Apartado 321, 38700 Santa Cruz de La Palma, Spain}

\altaffiltext{$\dagger$}{Email: xinfeng@vt.edu}


\begin{abstract}

We conducted a survey of quasar outflows using the VLT/Xshooter spectrograph.  When choosing the 14 BAL and mini-BALs comprising this sample, the data did not cover the  \siv\ and \siv* troughs, whose ratio can be used to determine the distance of the outflows from the central source (R). Therefore, this 
``Blind Survey'' is unbiased towards a particular distance scale.  Out of the eight outflows where R can be measured, six have 
$R > 100$ pc (spanning the range 100--4500 pc), one has $R > 10$ pc, and only one (at $R < 60$ pc) is compatible with a much smaller R scale. At least two of the outflows have a kinetic luminosity  greater than $0.5\%$ of their Eddington luminosity, implying that they are able to provide significant AGN feedback. The outflows span a range of $0$ to $-10000$ km s$^{-1}$ in velocity; total column density between $10^{20}-10^{22.5}$ cm$^{-2}$; ionization parameter (\Uh) in the range 0.01--1; and electron number density between $10^{3}-10^{5.5}$ cm$^{-3}$, with one upper and one lower limit.
The results of this survey can be extrapolated to the majority of BAL outflows, implying that most of these outflows are situated far away from the AGN accretion disk; and that a significant portion of them can contribute to AGN feedback processes.
\end{abstract}


\section{INTRODUCTION}

Quasar outflows, observed as blueshifted absorption features in spectra, are important candidates for active galactic nucleus (AGN) feedback. The energy and momentum injected by AGN outflows to the galactic host are invoked to explain the co-evolution of supermassive black holes (SMBH) with their respective host galaxy \citep{Silk98,Ferrarese00,Di05,Ostriker10, Hopkins10, Soker11, Hopkins16,Ciotti17,Angles17}, the shape of the observed quasar luminosity function \citep{Hopkins05a,Hopkins07,Hopkins10b,FaucherGiguere12}, and the metal enrichment of galaxy clusters \citep{Moll07,McCarthy10,Baskin12,tay15}.

One of the most important parameters needed to assess the contribution of an outflow to AGN feedback is the distance (R) of the outflow from the central source. In some theoretical models, the outflows originate from accretion disk winds \citep{Murray95,Proga00,Proga04} at $\sim 10^{16}\ $cm from the central source, assuming a $10^8M_{\bigodot}$ black hole. The most straight forward method for determinating R is based on spectral imaging using an Integral Field Unit (IFU) \citep{Riffel11,Rupke11,Harrison12, Riffel13,  Liu13a,Liu13b,Liu14,Liu15,Harrison14, Diniz15, Fischer15, McElroy15}. However, current instruments cannot resolve outflows on scales smaller than a few kpc in high luminosity quasars.

Our group and others have focused on determining R by using troughs from excited ionic states observed in broad absorption line (BAL) and mini-BAL outflows. The column density ratio between the excited and ground states yields the outflow's electron number density (\ne). Combined with the definition of the ionization parameter, R can be determined [see equation (1)]. There are more than 20 AGN outflows published using this method  \citep{Hamann01,deKool01,deKool02a,deKool02b,Gabel05,Moe09,Bautista10,Dunn10a,Aoki11,Arav12,Borguet12a,Borguet12b,Borguet13,Edmonds11,Arav13,Lucy14,Finn14,Chamberlain15a,Chamberlain15b,Arav18,Xu18}. Most of these measurements constrain the outflow to distances in the range of parsecs to several kilo-parsecs, which is much larger than the predictions from accretion disk wind models. Analyzing trough variability is another method to estimate R, but the results are model-dependent [see discussion in Section 7.1 of \cite{Arav18}].

About $\sim$90\% of all BAL outflows show absorption troughs only from high ionization species such as \civ\ and \siiv\ \citep{Trump06}. In order to determine R for these so-called HiBAL outflows, one needs    to measure troughs from excited states of high ionization species. 
As described in \cite{Arav18}, \siv\ (1062.66\angstrom) and \siv*\ (1072.96\angstrom) are the main transitions from highly ionized species, observable with ground-based observatories, for which \ne\ can be determined.

Most of these distance determinations arise from analysis of single objects. This can lead to selection biases by choosing the ``desired" object with certain properties. This will, in turn, cause a bias in the R determinations \citep[][but see Arav et al. 2018 Section 7.1]{Lucy14}. Therefore, it is hard to generalize these findings to the whole population of quasar outflows. 

One of the possible solutions to this problem is to conduct a ``Blind Survey" to get an unbiased sample. This means selecting targets where it is not known a priori if a certain density ratio is present, nor its probable value. Therefore, any selection bias due to choosing targets because of the ratio is eliminated.

In \cite{Arav18}, we followed this idea and introduced a ``Blind Survey" of 14 SDSS/BOSS objects, for which we then obtained observations from the Very Large Telescope (VLT). We discussed this survey and did a preliminary analysis which only used the depth ratio of \siv*/\siv\ to constrain R. From this survey, we have already published R determinations on three objects \citep{Borguet13, Chamberlain15b, Xu18}. This paper continues the work to complete the full analysis of the survey.

The structure of this paper is as follows.  In Section 2, we discuss the distance determination method. In Section 3, we present the details of our ``Blind Survey", including the selection criteria. In Section 4, we show the important parts of each spectra and describe the template fitting technique used to get column densities of the absorption troughs. The photoionization and density analyses are in Section 5 and Section 6, respectively. In Section 7, we present the distance and energetics for the outflows as well as discuss their implications in Section 8. Lastly, we summarize our results in Section 9.

\begin{figure}[hb]
\centering
\includegraphics[width=1.000\linewidth,keepaspectratio,trim={1cm 0cm 1cm 1cm}]{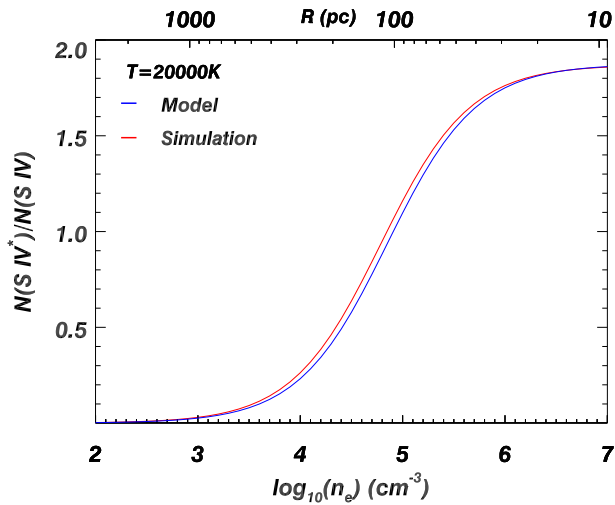} \\
\caption{ Column density ratio of N(\siv*)/N(\siv)\ versus \ne. The two curves are the comparison between the analytical formula [equation (\ref{Eq:Collisional})] and CHIANTI simulations. The difference is negligible. The top x-axis is the distance calculated from log(\ne) following equation (\ref{Eq:ionPoten}) using \Qh\ and \Uh\ derived for quasar J0941+1331 (see Section \ref{Energy}).}
\label{fig:SIVmod}
\end{figure}

\section{Distance Determination Method: \siv\ and \siv* }


Ionization equilibrium in an AGN outflow is dominated by photoionization \citep{Arav01}. A photoionized plasma is characterized by the ionization parameter, \Uh, where

\begin{equation}
\label{Eq:ionPoten}
\Uh=\frac{\Qh}{4\pi R^2 \nh c}
\end{equation}
$\Qh$ is the source emission rate of hydrogen ionizing photons, R is the distance of the outflow to the central source, $\nh$\ is the hydrogen number density, and c is the speed of light. 

In order to solve equation (\ref{Eq:ionPoten}) for R, we determine: a) \Uh\ from the photoionization modelling discussed in Section \ref{section:photo}; b) \Qh\ by integrating the spectral energy distribution (SED) for photon energies higher than the hydrogen ionization threshold; c) \nh\ from the relation between \nh\ and the electron number density (\ne): for a highly ionized plasma, $\nh \simeq 0.8\ne$. To determine \ne, we used troughs from ground- and excited-states of \siv\  \citep{Borguet12b,Arav18,Xu18}. The \siv*\ energy level is mainly populated by collisionally excited electrons from the ground-state of \siv\ \citep{Leighly09}. Therefore, the ratio between the column densities of \siv*\ and \siv\ can be used to determine \ne. Theoretically, from \cite{Arav18}, \ne\ is related to the column density ratio, N(\siv)/N(\siv*), by the expression: 

\begin{equation}
\label{Eq:Collisional}
n_e \simeq n_{cr}\left [\frac{2N(S\  IV)}{N(S\ IV^{*})}e^{-\Delta E/kT}-1 \right ]^{-1}
\end{equation}

where $n_{cr}=6.3\times10^4$ cm$^{-3}$ is the critical density for N(\siv)/N(\siv*)\ at T=20,000 K, N(\siv) and N(\siv*) are the column densities of the ground- and excited-states, respectively, $\Delta$E = 0.112 eV is the energy difference between these two states, T is the temperature, and k is the Boltzmann constant. In practice, we run grids of collisional excitation models using CHIANTI version 7.1.3 \citep{Landi13} to generate the N(\siv*)/N(\siv)\ ratio curve. We compare our measured ratio to the CHIANTI prediction to determine \ne. We show in figure \ref{fig:SIVmod} the comparison between equation (\ref{Eq:Collisional}) and the CHIANTI simulations. The difference is negligible. The measurements of \ne\ for our outflows using this method are discussed in Section \ref{section:siv}.

We note that there are other high ionization transitions near the wavelength positions of \siv\ and \siv*\ that can be used as well. For example, \niii\ (989.80\angstrom) and \niii*\ (991.57\angstrom) have similar ionization potentials to \siv\ and \siv*\ and can trace HiBALs. However, the troughs are hard to resolve since the velocity seperation between the \niii\ and \niii*\ transitions ($\sim$ 500 km s$^{-1}$) is much smaller than the seperation for the \siv\ and \siv*\ ones ($\sim$ 2900 km s$^{-1}$).

\section{Observation and Data Reduction}
\subsection{Blind Survey Setup}
Our group has published several R determinations using measurements of the N(\siv*)/N(\siv) ratio \citep[e.g.,][]{Borguet13,Chamberlain15b,Arav18,Xu18}. However, as noted above, these single object observations can suffer from selection biases. In \cite{Arav18}, we used the depth ratio of the \siv*\ and \siv\ troughs to constrain R. A depth ratio $>$ 1 results in R $\lesssim$ 100 pc, and a depth ratio $<$ 1 leads to R $\gtrsim$ 100 pc. Therefore, knowing the depth ratio when choosing objects can lead to biased R determinations. 

In order to build a distribution of R from an unbiased sample, we conducted a ``Blind Survey". The sample was selected from the SDSS/BOSS quasar spectra catalog. SDSS covers the wavelength range between 3800\angstrom\ and 10,400\angstrom, while BOSS covers 3650\angstrom\ to 10,400\angstrom. For objects with redshifts $<$ 2.6, the \siv\ and \siv*\ troughs are not observable in this spectral range. Therefore, choosing SDSS and BOSS quasars with z $<$ 2.6 guarantee that we would not have a bias towards specific depth ratios of the \siv*\ and \siv\ troughs, hence the term ``Blind Survey". Two criteria were used to select objects from the SDSS/BOSS data: 1) bright objects (r band magnitude $\lesssim$ 18.8), in order to obtain a high signal to noise ratio in a reasonable exposure time; and 2) deep \siiv\ troughs, which increases the probability of detecting \siv\ and \siv* troughs. These two criteria also do not bias the selected objects to any particular ratio of N(\siv*)/N(\siv). 

In total, 14 objects were selected. Observations were taken with the VLT/X-Shooter spectrograph, where \siv\ and \siv*\ are in the range of 3000\angstrom\ to 3800\angstrom\ for SDSS objects, and between 3000\angstrom\ and 3650\angstrom\ for BOSS objects. Therefore, the sample is unbiased towards the ratio of N(\siv*)/N(\siv). 

\subsection{X-Shooter Sample}
Our results for this ``Blind\ Survey" are based on the analysis of high S/N, medium resolution (R $\sim$ 6000-9000) VLT/X-Shooter spectra from several VLT programs (PI: Benn, see table 1). X-Shooter is a second-generation spectrograph installed on VLT. In a single exposure, a wide spectral band width (3000 -- 25000\angstrom) is covered since, by design, the incoming light is split into three bands (UVB, VIS and NIR). All of these objects have wide \civ\ absorption lines, and are classified as either BAL quasars \citep[\civ\ width $ > $ 2000 km s$^{-1}$, see ][]{Weymann91}, or mini-BAL quasars \citep[2000 $>$ \civ\ width $>$ 500 km s$^{-1}$, see ][]{Hamann04}. Specifics of the observations for each object are shown in table 1.


\begin{deluxetable}{l c c c r }[ht]

\setlength{\tabcolsep}{0.02in} 
\tablecaption{ VLT/X-Shooter Blind Survey: Observation Details}
\tablehead{
\colhead{RA, Dec}  & \colhead{Proposal ID} & \colhead{$z$} & \colhead{$r$} & \colhead{Exp. }\\ 
\colhead{(1)} & \colhead{(2)} & \colhead{(3)} & \colhead{(4)} & \colhead{(5)}
}

\startdata
\multicolumn{5}{l}{Objects with identified \siv\ troughs:}\\
00:46:13.54$\;+$01:04:25.8  & 	091.B-0324(B)	&2.149 & 18.04  & 10.8\\
08:25:25.07$\;+$07:40:14.3  & 	092.B-0267(A)	&2.204 & 17.89  & 18.0\\
$^{a}$08:31:26.16$\;+$03:54:08.1  & 	092.B-0267(A)	&2.076 & 18.27  & 10.8\\
09:41:11.12$\;+$13:31:31.2  & 	092.B-0267(A)	&2.021 & 18.15  & 10.8\\
$^{b}$11:11:10.15$\;+$14:37:57.1  & 	092.B-0267(A)	&2.138 & 18.03  & 10.8\\
11:35:12.70$\;+$16:15:50.7  & 	091.B-0324(A)	&2.004 & 18.36  & 7.2\\
$^{c}$15:12:49.29$\;+$11:19:29.4  & 	087.B-0229(A)	&2.109 & 17.65  & 8.4\\

\\ [2mm]
\multicolumn{5}{l}{Objects without identified \siv\ troughs:}\\
01:27:48.40$\;-$00:13:34.2  &	090.B-0424(A)	&2.076  & 18.10 & 6.4\\
02:52:21.20$\;-$08:55:16.5  &	090.B-0424(A)	&2.296  & 18.00 & 9.6\\
08:11:14.50$\;+$17:20:55.2  &	090.B-0424(B)	&2.329  & 18.00 & 12.8\\
08:22:24.90$\;+$08:02:52.5  &	090.B-0424(B)	&1.976  & 17.90 & 9.6\\
11:00:41.20$\;+$00:36:29.4  & 	087.B-0229(A)	&2.020  & 18.60 & 11.2\\
21:49:01.00$\;-$07:31:43.5  &	087.B-0229(B)	&2.211  & 18.10 & 8.4\\
22:18:38.30$\;-$08:34:51.2  & 	091.B-0324(B)	&2.522  & 18.80 & 9.6\\
\enddata
\tablecomments{
(1) Right ascension and declination. (2) VLT/X-Shooter Proposal ID (3) Redshifts of quasars. (4) $r$-band magnitude obtained through PSF fitting. (5) Total exposure time in kilo-seconds.
\textbf{References:} a: \cite{Chamberlain15a}; b: \cite{Xu18}; c: \cite{Borguet12b}.\\
 } 

\label{tab:VLT}
\end{deluxetable}

\begin{deluxetable}{r c c c }[ht]

\setlength{\tabcolsep}{0.02in} 
\tablecaption{ VLT/X-Shooter Blind Survey Objects Properties}
\tablehead{
\colhead{Object}  & \colhead{$^{a}$BH Mass} & \colhead{$log(L_{bol})$} & \colhead{$^{b}$\civ\ Abs. Width}\\ 
\colhead{} & \colhead{log(M/$M_{\bigodot}$)} & \colhead{erg s$^{-1}$} & \colhead{(km/s)} 
}

\startdata
J0046+0104 	&9.0	&47.1&	3700\\
$^{m}$J0825+0740 	&8.8	&46.8&	1600\\	
J0831+0354 	&8.8	&46.9&	5000\\	
J0941+1331 	&8.8	&46.8&	8000\\	
$^{m}$J1111+1437 	&8.8	&46.9&	1800\\	
J1135+1615 	&9.0	&47.2&  9500\\			
$^{c}$J1512+1119 &9.1	&47.5&	2000\\	
\enddata
\tablecomments{
($^{a}$): Based on the virial theorem, black hole mass is estimated using the luminosity and full width half maximum (FWHM) of the \civ\ emission line following \cite{Park13} equation (3).
($^{b}$): \civ\ absorption trough width is measured for continous absorption below the normalized flux I = 0.9.
($^{c}$): J1512+1119 has two outflows with \siv\ absorption troughs.
($^{m}$): classified as mini-BAL quasars.
}
\label{tab:VLT2}
\end{deluxetable}

\section{Spectral Fitting}

\subsection{Unabsorbed Emission Model and Outflows}
Generally, the unabsorbed spectrum of AGN contains two components: 1) a continuum which is usually modelled by a power law and 2) emission lines that can be fit with Gaussian profiles. For all of the objects, we fit the continuum with these two components. 

In total, eight outflows from seven objects had measurable \siv\ absorption troughs. Six out of the eight are classified as BAL outflows, while the other two are classified as mini-BALs. Details of the outflows are given in table \ref{tab:VLT2}. \\

\begin{figure*}[htp]

\centering
	\includegraphics[angle=0,trim={2cm 0 0.cm 0},clip,width=0.40\linewidth,keepaspectratio]{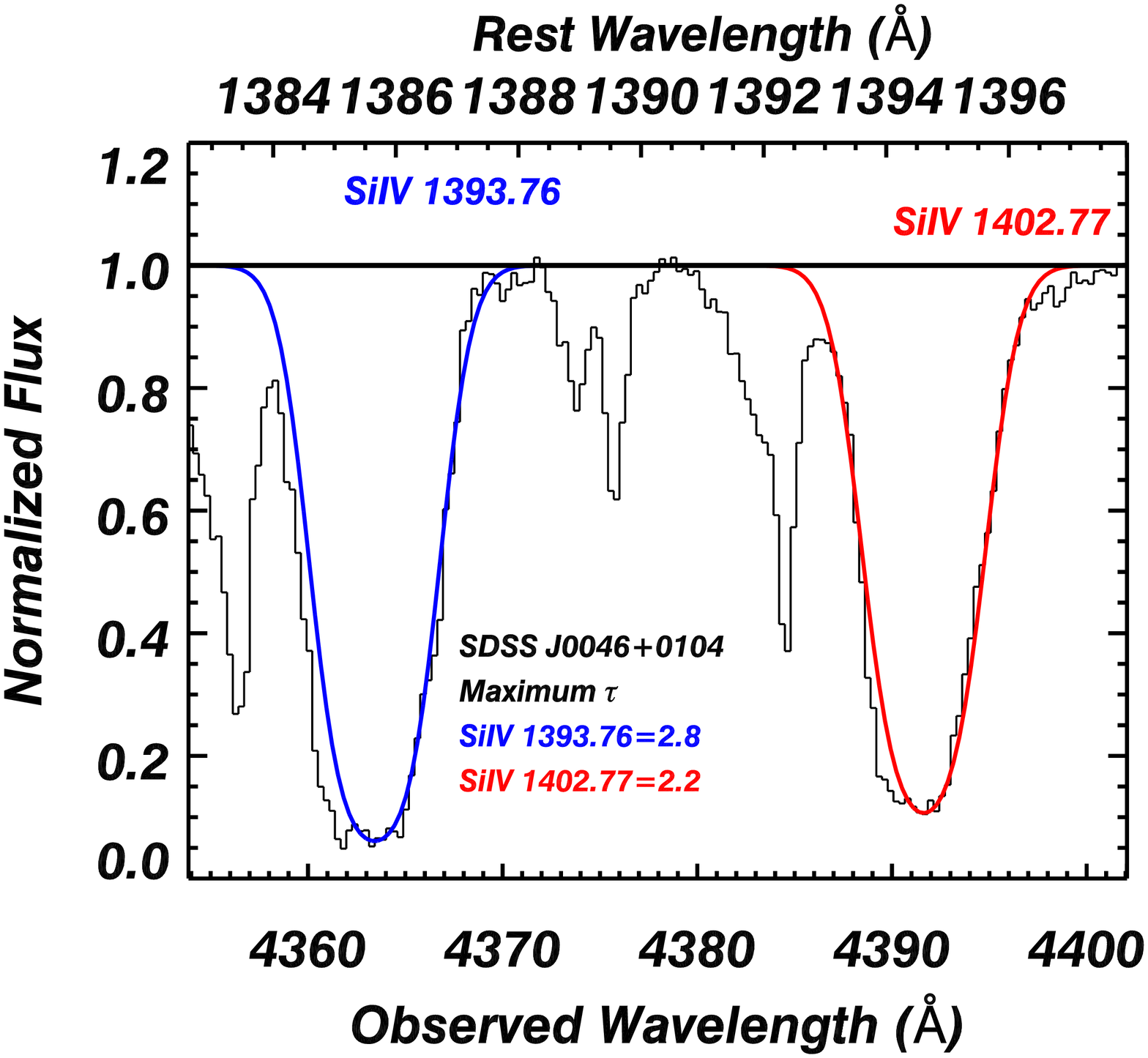}
	\includegraphics[angle=0,trim={2cm 0 0.cm 0},clip,width=0.40\linewidth,keepaspectratio]{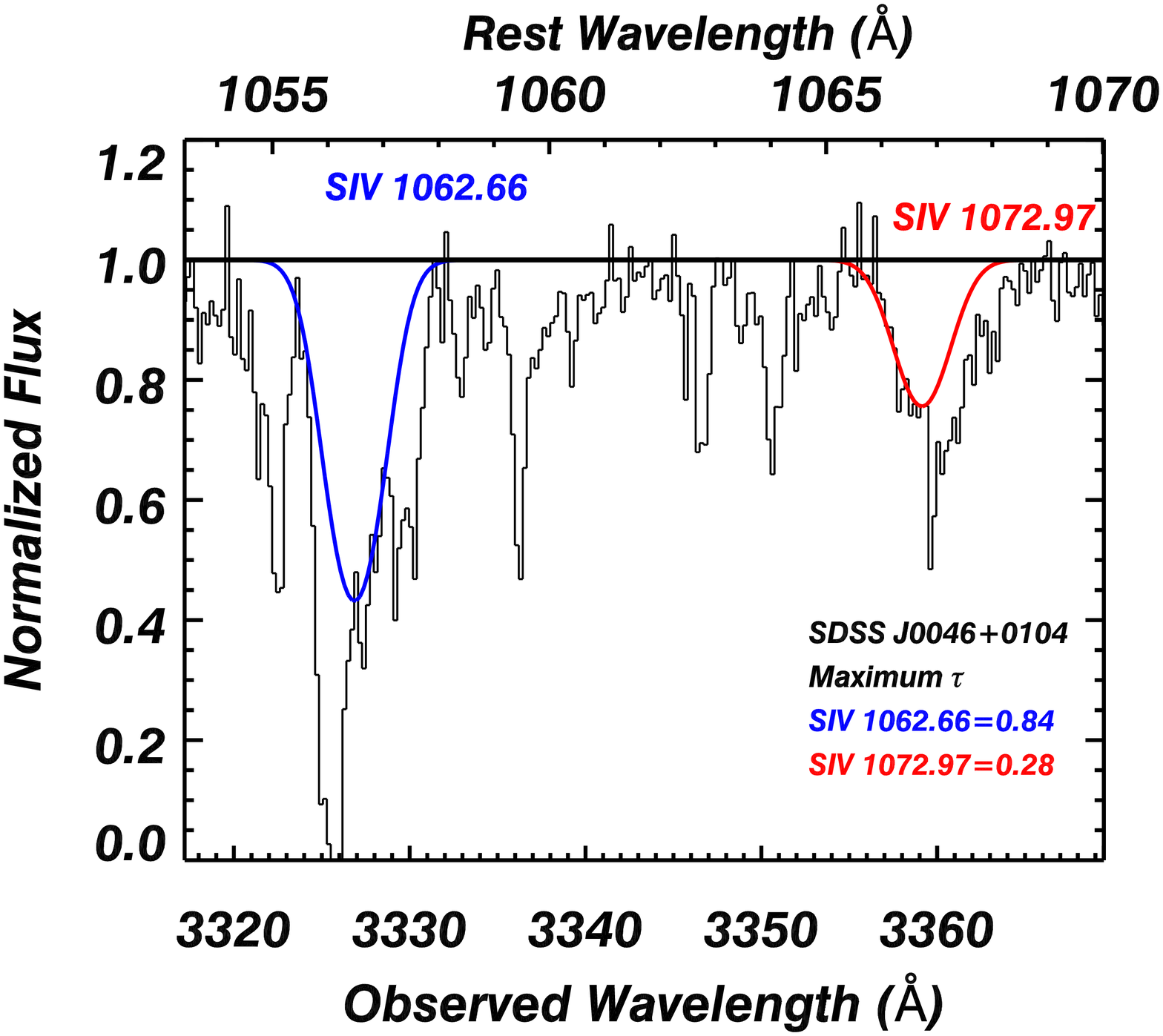}\\

	\includegraphics[angle=0,trim={2cm 0 0.cm 0},clip,width=0.40\linewidth,keepaspectratio]{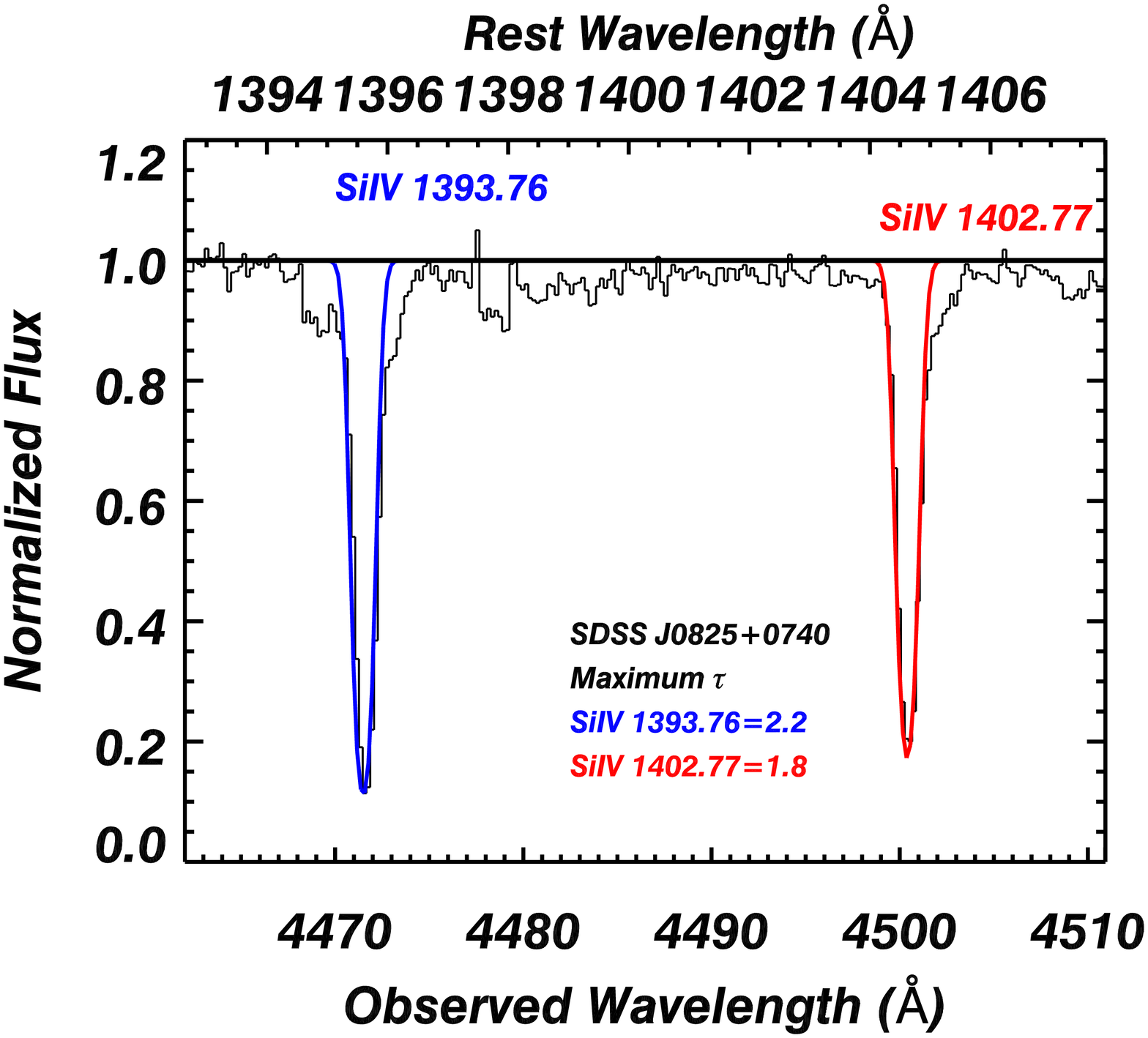}
	\includegraphics[angle=0,trim={2cm 0 0.cm 0},clip,width=0.40\linewidth,keepaspectratio]{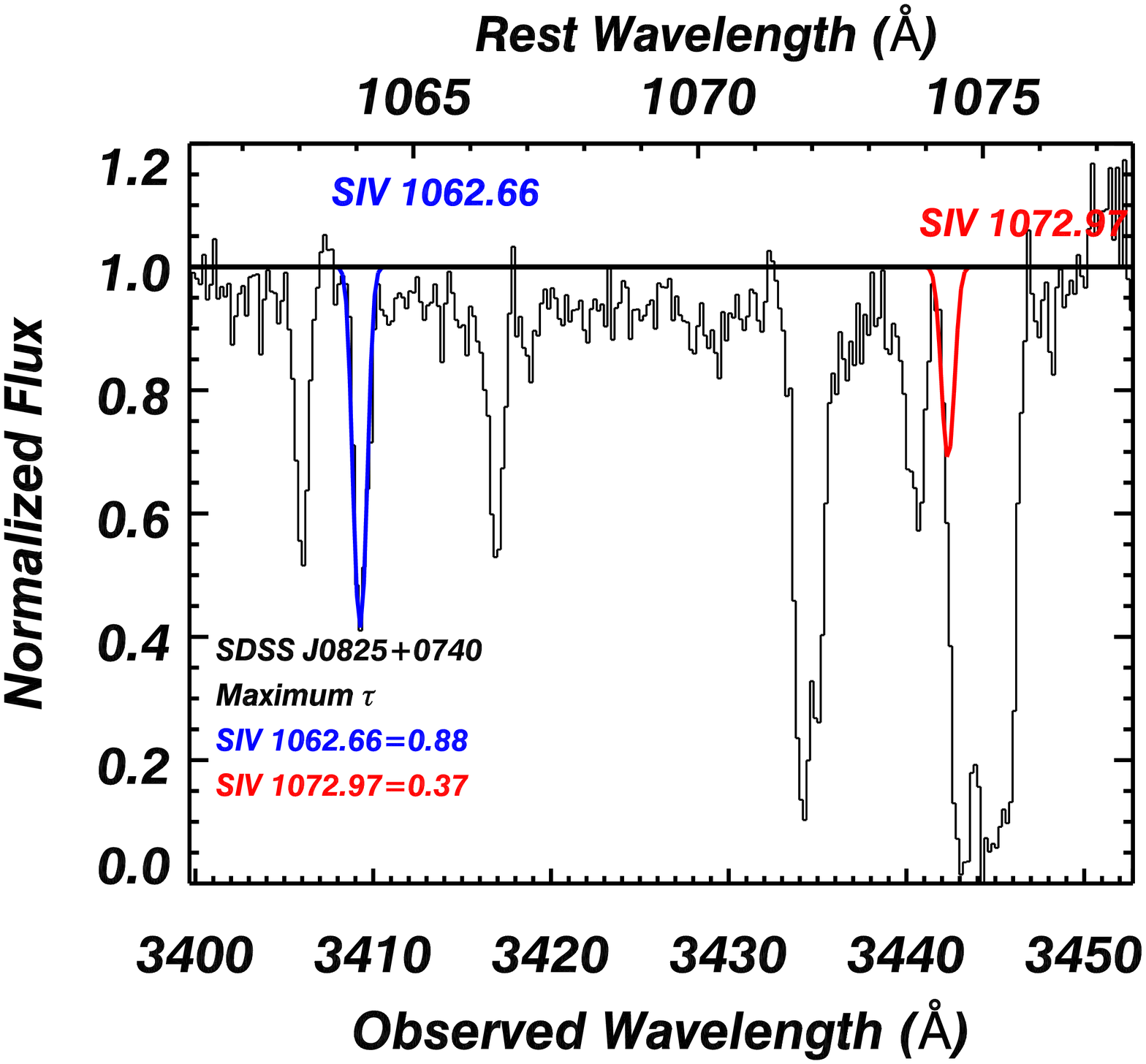}\\

	\includegraphics[angle=0,trim={2cm 0 0.cm 0},clip,width=0.40\linewidth,keepaspectratio]{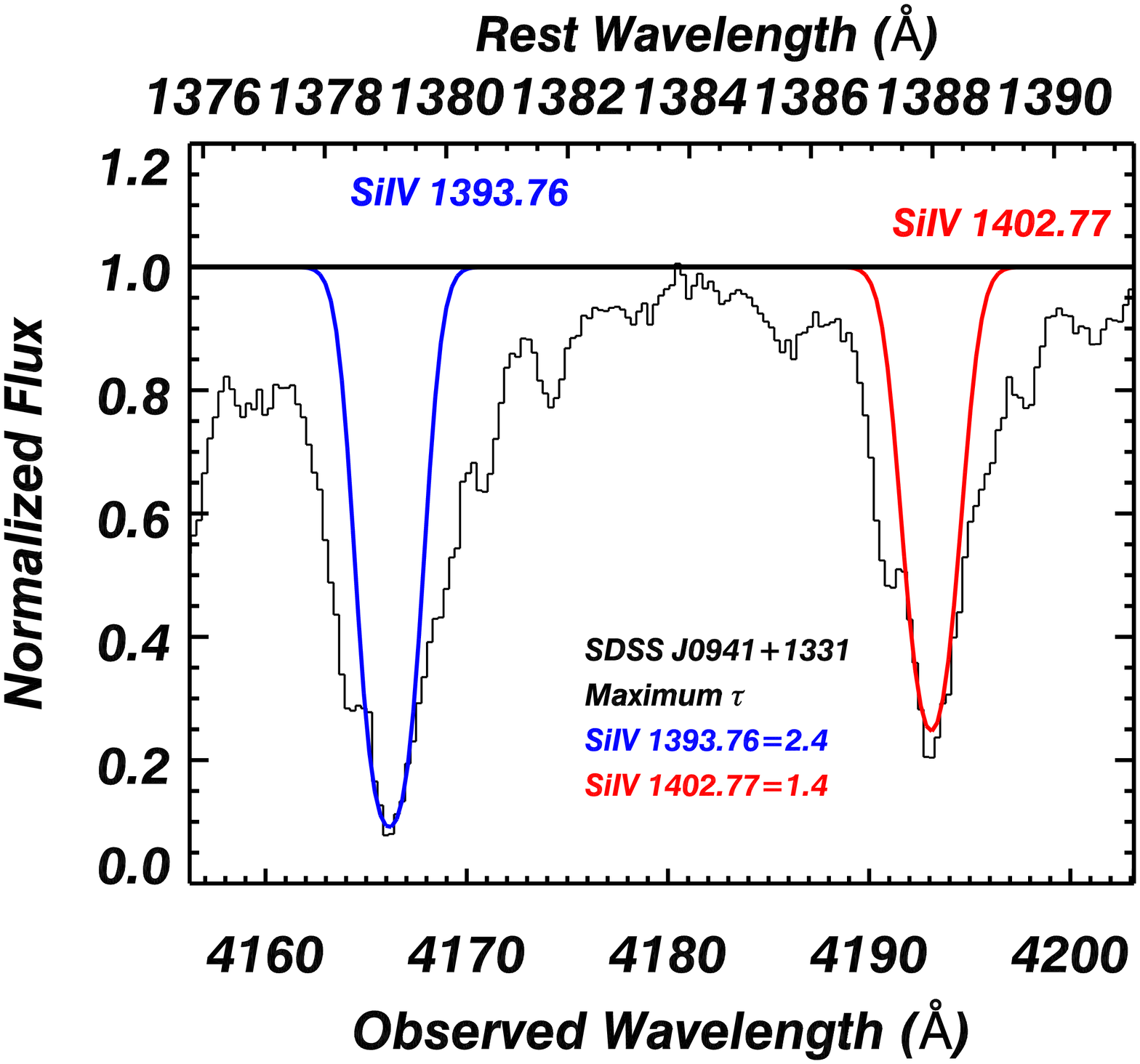}
	\includegraphics[angle=0,trim={2cm 0 0.cm 0},clip,width=0.40\linewidth,keepaspectratio]{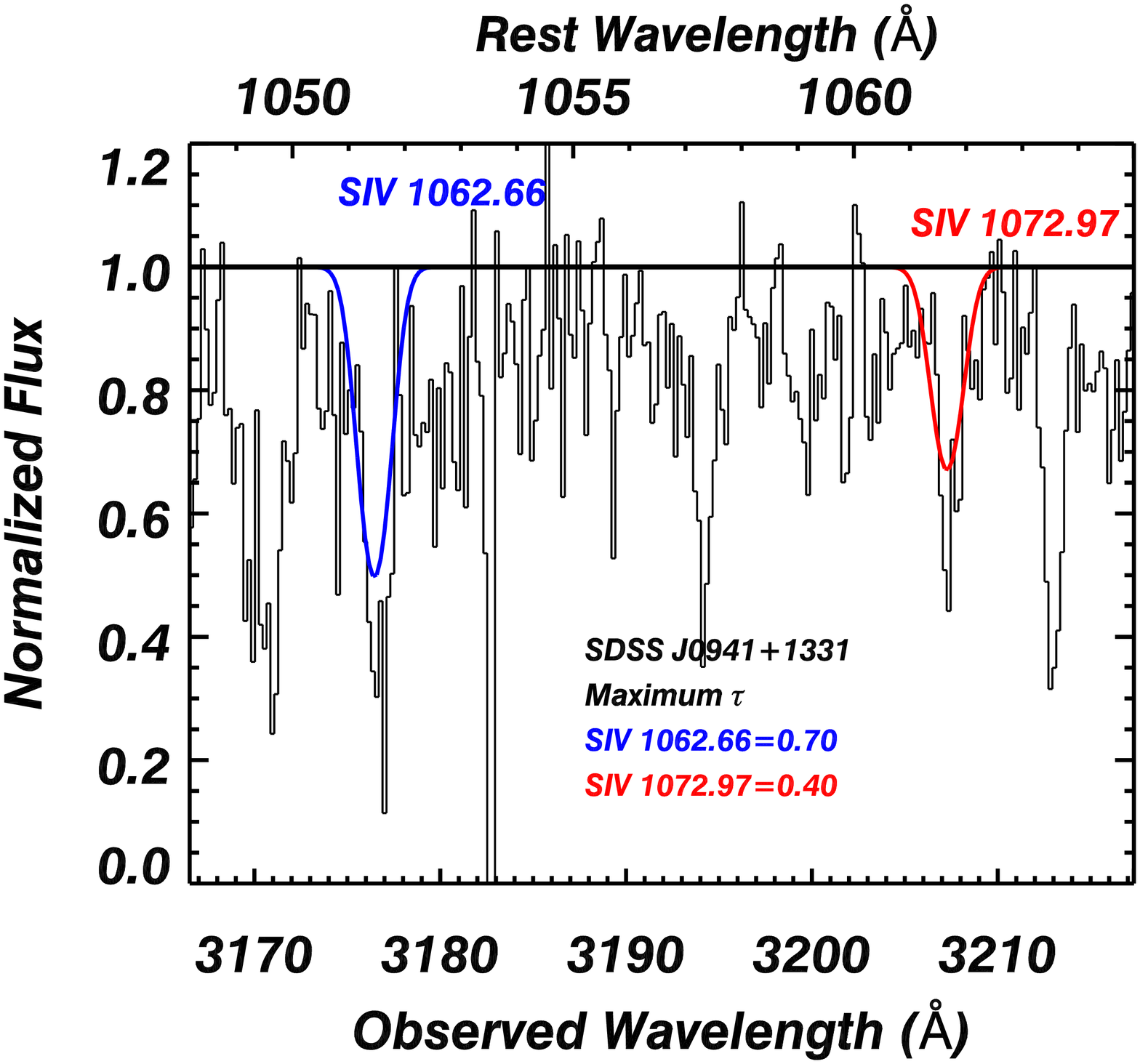}\\

	\includegraphics[angle=0,trim={2cm 0 0.cm 0},clip,width=0.40\linewidth,keepaspectratio]{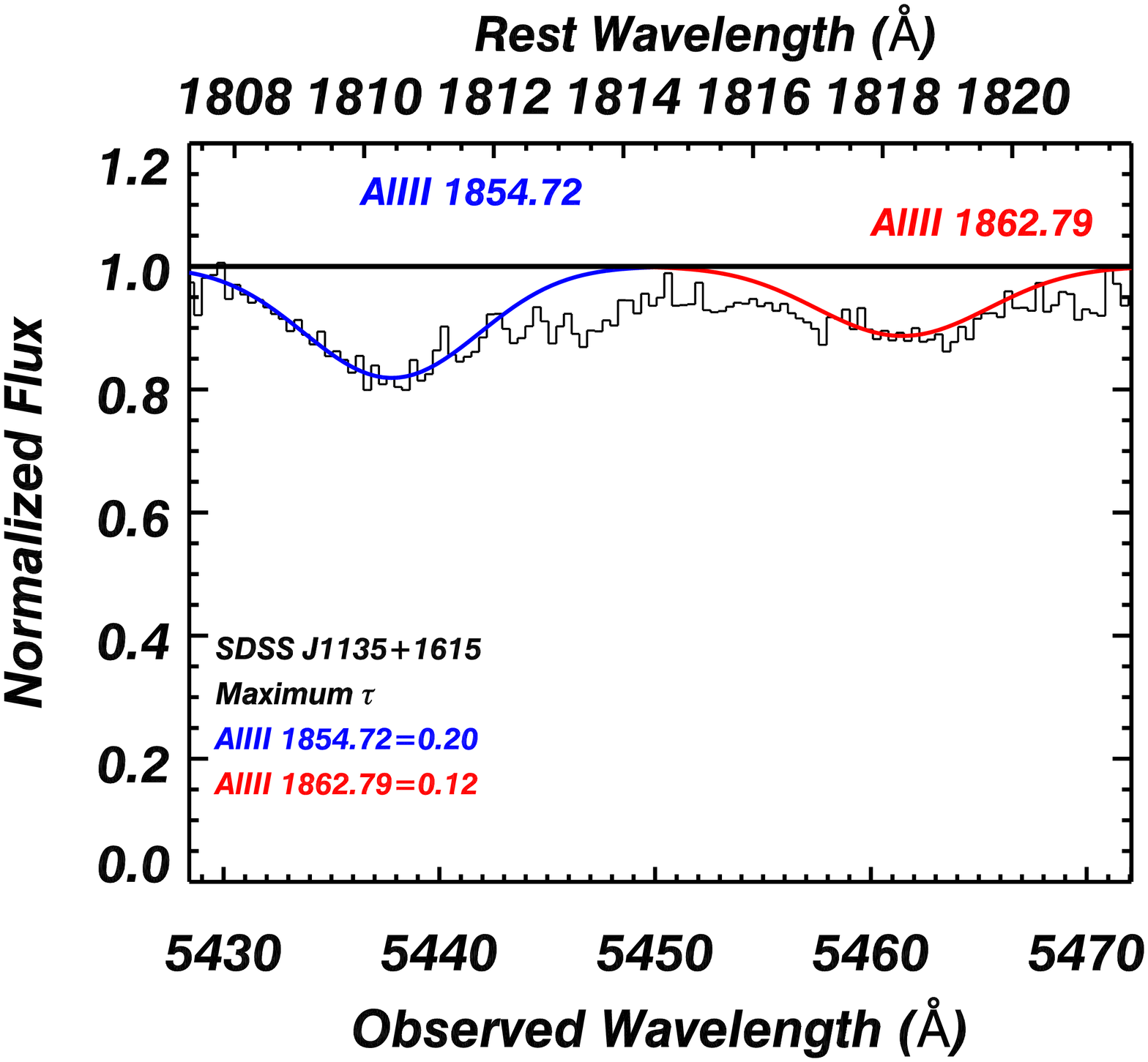}
	\includegraphics[angle=0,trim={2cm 0 0.cm 0},clip,width=0.40\linewidth,keepaspectratio]{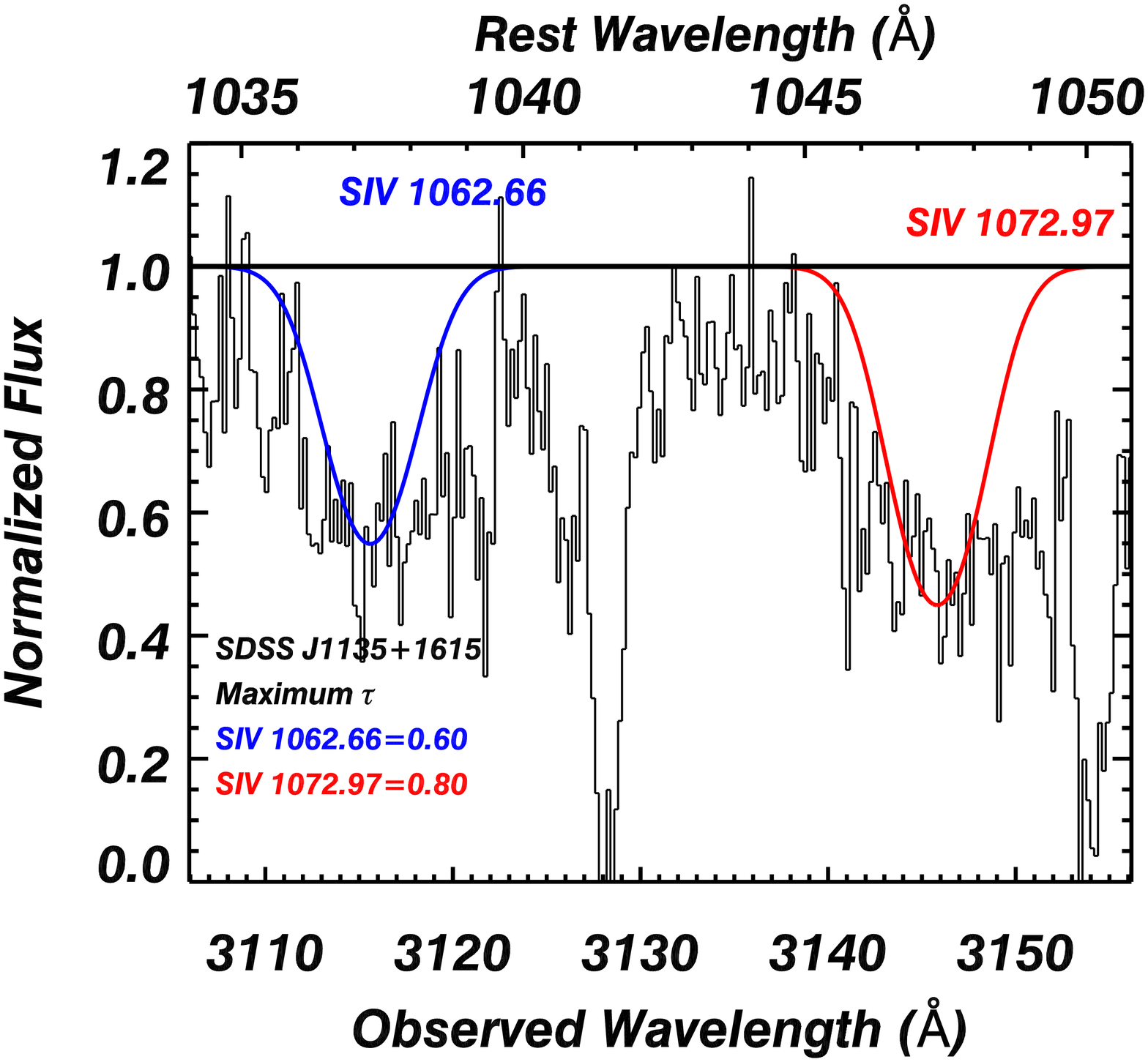}\\

\caption{Each row shows the \siiv\ and \siv\ template fitting of one object in our survey. Other objects have already been published: J0831+0354 \citep{Chamberlain15a}, J1512+1119 \citep{Borguet12b} and J1111+1437\citep{Xu18}}
\label{fig:fit}
\end{figure*}

\subsection{Template Fitting and Column Density Extraction}
\label{sec:template}

To measure ionic column densities from the observed troughs, we use the template fitting process described in \cite{Xu18}. The unblended \siiv\ 1402.77\angstrom\ troughs (for one case we used \aliii\ 1854.72\angstrom, see section 4.3) were modelled by a Gaussian optical depth profile. Blended and contaminated troughs (including \siv\ and \siv*) can then be fitted with this Gaussian template by scaling the Gaussian's depth while keeping the velocity centroid and width unchanged. Four outflows have been published previously \citep{Borguet12b, Borguet13,Chamberlain15b, Xu18}. We show the fitting results for the other four outflows in figure \ref{fig:fit}.\\   

\subsection{Notes on Individual Objects}

SDSS J0046+0104 has a wide \siiv\ absorption trough, with narrower ones on the left wing from other outflows. For the \siv*/\siv\ region, contamination by the Ly$\alpha$ forest is present on the right wing of \siv*\ and both wings of \siv. Using the \siiv\ Gaussian template, we were able to factor out the contamination and obtain reliable column density measurements.\\

SDSS J0825+0740 has one of the two mini-BAL outflows in our sample. The \siiv\ is quite narrow and without any contamination, as is the \siv\ absorption trough. However, the \siv*\ absorption trough is heavily contaminated by a strong, likely Ly$\alpha$, absorption trough from an intervening system. In this case, the template was fit to the left wing of the \siv*\ trough.\\

SDSS 0941+1331 has wide \siiv\ absorption troughs, but the center is fit well with a narrow Gaussian model. The narrow Gaussian of the \siiv\ template fits the \siv\ and \siv* absorption troughs well and removes the contamination in their centers.\\

For SDSS J1135+1615, the \siiv\ absorption troughs are saturated and flat. Therefore, we used \aliii\ 1854.72\angstrom\ as the template, since it is not blended with other troughs. Fitting the \siv\ and \siv*\ absorption troughs with this template shows that the \siv*\ absorption trough is actually deeper than the \siv\ absorption trough (see table \ref{tableCompare} for more details).

\section{Photoionization Analysis}
\label{section:photo}
In order to derive the physical conditions of the outflow, we run the spectral synthesis code Cloudy (version c17.00, \cite{Ferland17}) to generate a grid of photoionization simulations. Cloudy self-consistently solves the photoionization and thermal equations necessary to model the physical conditions in the gas. The absorbers are modelled as plane-parallel, thin slabs of constant density gas with solar metallicity. They are irradiated by the flux from the UV-soft SED \citep{Dunn10a}. To make the grid, we varied the ionization parameter such that log(\Uh) covers the range between -5.0 and 3.0 in steps of 0.05 dex. At each \Uh, Cloudy calculates the predicted column densities of each ion for a range of hydrogen column densities, \Nh. The largest value of \Nh\ is determined when the proton density reaches 10\% of the total hydrogen density. The predicted column densities are compared to the measured values of each ion. The best fit solution is found through $\chi^2$-minimization of the difference between the model prediction and the measured column densities. 

\begin{figure}[hb]
\centering
\includegraphics[width=0.96\linewidth,keepaspectratio,trim={1.5cm 0cm 0cm 1.5cm}]{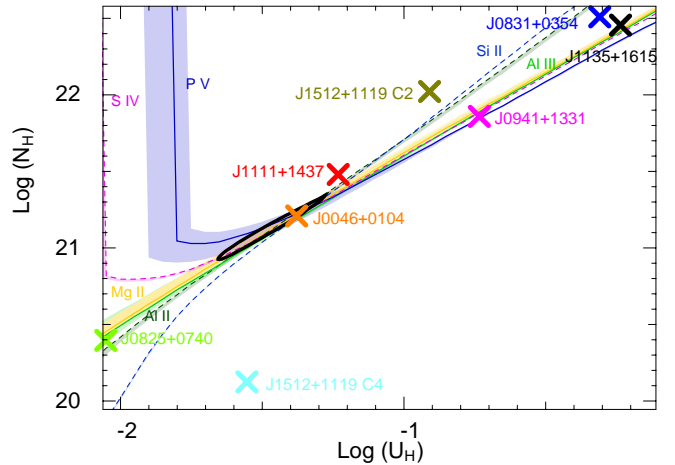} \\
\caption{The photoionization solutions for the eight outflows are shown. The colored contours represent the models whose column densities are consistent with the observed values for J0046+0104. The $\chi^2-$minimization solution is indicated by the orange cross and the 1$\sigma$ confidence level by the black ellipse. The solid lines represent measurements while the dashed lines represent lower limits.}
\label{fig:photo}
\end{figure}

We show these solutions as crosses for all eight outflows in figure \ref{fig:photo}. The color contours are for the regions where the model predicted column densities are consistent with the measured column densities for SDSS J0046+0104. The solutions spread over 2 dex in both log(\Uh) and log(\Nh), similar to the range of \Uh\ and \Nh\ of previously published outflows \citep[See table 2 in][]{Xu18}. 

\section{The density-sensitive troughs: \siv \ and \siv*}
\label{section:siv}
\begin{figure}[hb]
\centering
\includegraphics[width=1.04\linewidth,keepaspectratio,trim={4cm 1.5cm 0cm 0cm}]{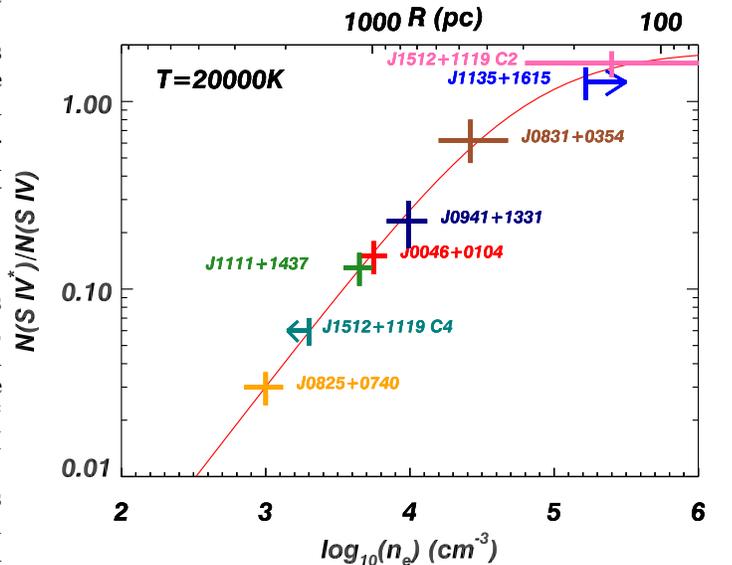}
\vspace*{5mm}
\caption{The column density ratio of N(\siv*)/N(\siv) versus log(\ne). The crosses and arrows are  solutions for different outflows. The top x-axis shows the conversion of $n_e$ to the outflow distance from the central source in the case of SDSS J0046+0104.}
\label{fig:SIV}
\end{figure}

As explained in section 2, the \siv*\  energy level is populated by collisional excitations from free electrons, so the ratio between the column densities of \siv*\ and \siv\ can be used as a diagnostic for \ne. In figure \ref{fig:SIV}, we show the ratio between the column densities of \siv* and \siv\ versus \ne. The curve is produced by CHIANTI version 7.1.3 assuming a temperature of 20,000 K. The temperature of the \siv\ zone in each outflow is taken from the best fit Cloudy simulation. The full range of temperatures is from 8000 to 20,000 K, which corresponds to a change in log(\ne) of less than 0.15 dex. We take the total N(\siv) from the photoionization model, and follow the procedure outlined in Section 5 of \cite{Xu18} to determine the N(\siv*)/N(\siv) ratio. Each colored cross in figure \ref{fig:SIV} represents the solution for one outflow, where the errors are represented by the vertical and horizontal bars. The left and right arrows represent the upper and lower limits for J1512+1119 C4 and J1135+1615, respectively.\\ 

For SDSS J0825+0740, the \siv*\ trough is contaminated by absorption troughs from the \Lya\ forest, and the only information we can get is that N(\siv*)/N(\siv) $<$ 0.5. However, in this outflow we detect \Siii*\ (1264.74\angstrom) and \Siii\ (1260.44\angstrom) absorption troughs. Therefore, we used the N(\Siii*)/N(\Siii) ratio to measure \ne: log(\ne)=$3.0^{+0.12}_{-0.15}$ cm$^{-3}$. The corresponding N(\siv*)/N(\siv) value is shown in figure 4.

\begin{deluxetable*}{ l r l l l r l l l}[htb!]
\tablewidth{\textwidth}
\tabletypesize{\small}
\setlength{\tabcolsep}{0.02in}
\tablecaption{Xshooter Survey: Outflow Parameters\label{tab:compareTable}}
\tablehead{
 \colhead{Object}  & \colhead{ v} & \colhead{log(\Uh)} & \colhead{log(\Nh)} & \colhead{log(\ne)} & \colhead{R} & \colhead{$\dot{M}$} & \colhead{Log $\dot{E_{k}}$} & \colhead{ $\dot{E_{k}}/L_{Edd}$}
\\
\\ [-2mm]
 \colhead{}   & \colhead{(km s$^{-1}$)} & \colhead{dex}& \colhead{log(cm$^{-2}$)} & \colhead{log(cm$^{-3}$)}& \colhead{pc} & \colhead{($M_{\odot}$ yr$^{-1}$)} & \colhead{log(erg s$^{-1}$)} & \colhead{$\%$} 
}

\startdata

J0046+0104 	&-1730	&\textbf{-1.3}$^{+0.2}_{-0.4}$&\textbf{21.2}$^{+0.2}_{-0.4}$	&\textbf{3.8}$^{+0.1}_{-0.1}$ 	&\textbf{1000}$^{+310}_{-370}$	&\textbf{35}$^{+6}_{-9}$	&\textbf{43.5}$^{+0.07}_{-0.1}$		&\textbf{0.029}$^{+0.005}_{-0.007}$ \\
J0825+0740 	&+395	&\textbf{-2.1}$^{+0.3}_{-0.2}$&\textbf{20.4}$^{+0.4}_{-0.3}$	&\textbf{2.9}$^{+0.1}_{-0.1}$ 	&\textbf{4500}$^{+1000}_{-2100}$&\textbf{5.6}$^{+2.8}_{-1.5}$	&\textbf{41.4}$^{+0.18}_{-0.13}$	&$<$\textbf{0.0004} \\
J0831+0354	&-10800	&\textbf{-0.3}$^{+0.5}_{-0.5}$&\textbf{22.5}$^{+0.5}_{-0.4}$	&\textbf{4.4}$^{+0.3}_{-0.2}$	&\textbf{110}$^{+30}_{-25}$	&\textbf{410}$^{+530}_{-220}$	&\textbf{46.2}$^{+0.4}_{-0.3}$		&\textbf{14}$^{+18}_{-7.7}$\\	
J0941+1331 	&-3180	&\textbf{-0.7}$^{+0.3}_{-0.2}$&\textbf{21.9}$^{+0.3}_{-0.2}$	&\textbf{4.0}$^{+0.1}_{-0.1}$ 	&\textbf{290}$^{+60}_{-130}$	&\textbf{80}$^{+24}_{-5}$	&\textbf{44.4}$^{+0.11}_{-0.03}$	&\textbf{0.5}$^{+0.14}_{-0.04}$ \\
J1111+1437	&-1860 	&\textbf{-1.2}$^{+0.2}_{-0.2}$&\textbf{21.5}$^{+0.2}_{-0.3}$	&\textbf{3.6}$^{+0.1}_{-0.1}$	&\textbf{880}$^{+210}_{-260}$	&\textbf{55}$^{+10}_{-11}$	&\textbf{43.8}$^{+0.07}_{-0.1}$		&\textbf{0.03}$^{+0.005}_{-0.007}$\\	
J1135+1615	&-7250	&\textbf{-0.2}$^{+0.1}_{-0.2}$&\textbf{22.5}$^{+0.1}_{-0.2}$ 	&\textbf{$>$5.2}$^{+0.3}_{-0.2}$&\textbf{$<$60}$^{+18}_{-30}$	&\textbf{$<$150}$^{+7}_{-30}$	&\textbf{$<$45.4}$^{+0.02}_{-0.1}$	&$<\textbf{8}^{+0.4}_{-2.1}$ \\
J1512+1119,C4	&-1050	&\textbf{-1.5}$^{+0.2}_{-0.1}$&\textbf{20.1}$^{+0.3}_{-0.3}$	&$\leq \textbf{3.3}$ 		&$>\textbf{3000}$		&$>\textbf{3.4}$		&$>\textbf{42.1}$			&$>\textbf{0.0008}$	 \\
J1512+1119,C2	&-1850	&\textbf{-0.9}$^{+0.1}_{-0.1}$&\textbf{21.9}$^{+0.1}_{-0.1}$	&\textbf{5.4}$^{+2.7}_{-0.6}$ 	&\textbf{10 -- 300}	&\textbf{1 -- 55}		&\textbf{43}$^{+0.8}_{-0.9}$		&$<\textbf{0.04}$\\


\vspace{-2.2mm}
\enddata

\tablecomments{
\\
(1). All solutions assumed the UVsoft SED and metallicity Z = $Z_{\bigodot}$.\\
\indent\textbf{References:} SDSS J0831+0354: \cite{Chamberlain15a}; SDSS J1111+1437: \cite{Xu18}; SDSS J1512+1119: \cite{Borguet12b}.\\
}
\label{tableCompare}
\end{deluxetable*}

\section{Outflow Distance, Mass Flux, and Energetics}
\label{Energy}
In Sections 5 and 6, we determined \Uh\ and \ne\ for each outflow (recall that $\nh \simeq 0.8\ne$ for highly ionized plasma). All that remains to solve equation (\ref{Eq:ionPoten}) for R is to determine $\Qh$. $\Qh$ is calculated by integrating the UV-soft SED for energies above the ionization threshold of hydrogen, 13.6 eV [we adopt an h = 0.696, $\Omega_{m}$ = 0.286, and $\Omega_{\Lambda} = 0.714$ cosmology \citep{Bennett14}]. We show the distances in table \ref{tableCompare}. We find that six out of the eight outflows have R $>$ 100 pc, one of the outflows has R $>$ 10 pc (J1512+1119, C2), and the only outflow that could be closer to the central source is J1135+1615 (R $<$ 60 pc).

Assuming the outflow is in the form of a thin partial shell, its mass flow rate ($\dot{M}$) and kinetic luminosity ($\dot{E_{k}}$) are given by \citep{Borguet12a}:
\begin{equation}\label{eq:1}
\begin{split}
\dot{M}\simeq 4\pi \Omega R\Nh \mu m_p v 
\end{split}
\end{equation}

\begin{equation}\label{eq:2}
\begin{split}
\dot{E}_{k}\simeq \frac{1}{2} \dot{M}v^2
\end{split}
\end{equation}

where R is the distance of the outflow from the central source, $\Omega=0.08$ is the global covering factor for outflows showing \siv \ absorption troughs \citep{Borguet13}, $\mu$ = 1.4 is the mean atomic mass per proton, $m_p$ is the proton mass, \Nh\ is the absorber's total hydrogen column density, 
and $v$ is the radial velocity of the outflow. The results of $\dot{M}$ and $\dot{E_{k}}$ are shown in table \ref{tableCompare}. Since the errors for \Nh\ and \Uh\ are correlated (see the 1$\sigma$\ error ellipse in figure \ref{fig:photo}), and R is a function of \Uh, the errors in R and \Nh\ are also correlated. The error bars shown in the last three columns of table \ref{tableCompare} account for this correlation \citep[for more discussion, see ][]{Xu18}.

\section{Discussion}
\label{sec:discussion}
\subsection{Distance Determination Method and Results}
It is crucial to determine the distance of an AGN outflow to the central source in order to measure the influence the AGN has on its host galaxy. This paper focused on using density-sensitive transitions from \siv\ and \siv*\ to measure \ne, so that equation (\ref{Eq:ionPoten}) could be solved. There are a couple constraints to this method: 1) These UV absorption lines must be strong enough and not severely contaminated by $Ly\alpha$ forest lines; and 2) The redshift must be greater than two for ground-based observations. This method has also been successfully applied to several objects with other density diagnostic troughs, including N(\Feii*)/N(\Feii) \citep{Korista08, Aoki11, Lucy14}, N(\Siii*)/N(\Siii) \citep{Dunn10a,Moe09}, N(\niii*)/N(\niii) \citep{Chamberlain15a}, N(\siv*)/N(\siv) \citep{Borguet13,Chamberlain15b,Arav18,Xu18} and N(\oiv*)/N(\oiv) \citep{Arav13}.\\

As stated in the introduction, IFU spectroscopy is the most straight forward method to determine the distance of an AGN outflow. However, the limitation of ground-based IFU instruments is that the angular resolution is around $ 0.5''$. For a z = 2 quasar, this angular resolution translates to a spatial distance of four kpc. This does not probe for outflows existing at much smaller distances (0.01 -- 1000 pc). See Section 7.1 in \cite{Arav18} for a detailed discussion of all R determination methods.

\subsection {Physical Properties of the Sample Outflows}

From the observations and our analysis, important physical properties emerge: 1) All the objects have broad absorption troughs. Based on the \civ\ absorption troughs width, six BAL outflows and two mini-BAL outflows are included in our sample; 2) The log(\Nh) and log(\Uh) shown in figure \ref{fig:photo} spread around two dex in log space, covering both high and low column density objects. This sample includes not only highly ionized outflows with log(\Uh) near zero, but also less ionized outflows with log(\Uh) around -2; 3) From figure \ref{fig:SIV}, the derived log(\ne) span a range from 3 to around 5.5, with one outflow showing an upper limit and another outflow a lower limit; and 4) As shown in table \ref{tableCompare}, six out of the eight outflows have R $>$ 100 pc, one of them has R $>$ 10 pc, and one outflow has R $<$ 60 pc.

\subsection {Applicability of the ``Blind Survey" Results to the Broader BAL Population}

Our sample avoids the selection biases mentioned in Section 3.1. 
We found that 75\% of the \siv\ outflows (six out of eight) have R $>$ 100 pc. \cite{Arav18} showed that under plausible assumptions these results can be extrapolated to the entire population of HiBAL outflows. Our results (see table 3) confirm the assertion made in \cite{Arav18} that outflows with lower \Nh\ are not preferentially found as smaller distances than outflows with higher \Nh.



\subsection{Outflow Energetics and AGN Feedback}

AGN outflows are believed to have significant feedback on their environment that contributes to the co-evolution of SMBHs and their host galaxies \citep[for a review, see ][]{Kormendy13}.  If the initial feedback directly heats the galactic gas, 5\% of the radiated energy should come from the kinetic energy of the outflow to heat the intra-cluster medium (ICM). If the outflow follows a `two-stage' model, only $\sim$0.5 \% of the luminosity is needed to drive the initial hot outflow (Hopkins \& Elvis 2010). The Eddington luminosity is calculated from the mass of the BH. Comparing our results from table \ref{tableCompare} with the Eddington luminosity of each object, at least two objects have an $\dot{E}_{k}/L_{Edd}$ $\gtrsim $ 0.5 \%, which means they can be a significant contributor to AGN feedback. Another object (J1135+1615) can also have high enough $\dot{E}_{k}/L_{Edd}$ if its distance is close to the derived upper limit.


\section{Summary}

We presented detailed results from a ``Blind Survey" of 14 quasars, where half show \siv\ outflow troughs. The other half either didn't have detectable \siv\ absorption troughs associated with an outflow system, or there was heavy contamination and the existence of a \siv\ absorption trough couldn't be determined. The results are summarized as follows:\\

1. A total of eight \siv\ outflows were detected. \\

2. At least six out of the eight outflows have R $>$ 100 pc, one of them has R $>$ 10 pc (J1512+1119, C2), and the other one has R $<$ 60 pc (J1135+1615). Two of them have a large enough Eddington ratio for significant AGN feedback.\\

3. For each of these outflows, we constrain \ne. The \ne\ span a range of over 2.5 orders of magnitude.\\

4. The photoionization solutions of all the outflows span over 2 dex in both log(\Uh) and log(\Nh).\\

5. The higher quality data from our VLT/X-Shooter survey yields that 75\% of the BALQSO outflows have R $>$ 100 pc, compared to 50\% in \cite{Arav18}. However, the sample size is smaller at eight \siv\ outflows compared to 34 in the previous study.\\

6. As shown by \cite{Arav18}, these results can be extrapolated to 90\% of the BAL outflow population.

\acknowledgments

NA acknowledges support from NSF grant AST 1413319, as well
as NASA STScI grants GO 11686, 12022, 14242, 14054, and 14176, and NASA ADAP 48020.

Based on observations collected at the European Organisation for
Astronomical Research in the Southern Hemisphere
under ESO programmes: 087.B-0229(A), 090.B-0424(A), 091.B-0324(B), 092.B-0267(A)(PI: Benn).

CHIANTI is a collaborative project involving George Mason University, the University of Michigan (USA) and the University of Cambridge (UK).

\bibliography{apj-jour,dsr-refs}

\end{document}